\documentclass[epj]{svjourAdelaide}
\usepackage{graphicx}

\newcommand{\be}{\begin{equation}}
\newcommand{\ee}{\end{equation}}

\newcommand{\calO}{{\mathcal O}}

\begin{document}
\title{FLIC Fermions and Hadron Phenomenology}
\author{D.~B.~Leinweber\inst{1} 
\thanks{Plenary talk presented by Derek Leinweber.}%
\and J.~N.~Hedditch \inst{1}
\and W.~Melnitchouk \inst{1,2}
\and A.~W.~Thomas \inst{1}
\and A.~G.~Williams \inst{1}
\and R.~D.~Young \inst{1}
\and J.~M.~Zanotti \inst{1}
\and J.~B.~Zhang \inst{1}
}                     
%
%
\institute{Centre for the Subatomic Structure of Matter and Department
of Physics and Mathematical Physics, University of Adelaide, Adelaide,
SA 5005, Australia 
\and 
Jefferson Lab, 12000 Jefferson Avenue, Newport News, VA 23606, U.S.A. 
}
\date{Received: 27 September 2002}
%
\abstract{ 
A pedagogical overview of the formulation of the Fat Link Irrelevant
Clover (FLIC) fermion action and its associated phenomenology is
described.  The scaling analysis indicates FLIC fermions provide a new
form of nonperturbative ${\mathcal O}(a)$ improvement where
near-continuum results are obtained at finite lattice spacing.
Spin-1/2 and spin-3/2, even and odd parity baryon resonances are
investigated in quenched QCD, where the nature of the Roper resonance
and $\Lambda^*(1405)$ are of particular interest.  FLIC fermions allow
efficient access to the light quark-mass regime, where evidence of
chiral nonanalytic behavior in the $\Delta$ mass is observed.
} 
\maketitle
\section{FLIC Fermions}
\label{FLICfermions}

The CSSM lattice collaboration has been examining the merits of a new
lattice fermion action \cite{FATJAMES} in which the (irrelevant)
operators introduced to remove fermion doublers and lattice spacing
artifacts are constructed with smoothed links.  These links are
created via APE smearing \cite{ape}; a process that averages a link
with its nearest transverse neighbors in a gauge invariant manner.
Iteration of the averaging process generates a ``fat'' link.  The use
of links in which short-distance fluctuations have been removed
simplifies the determination of the coefficients of the improvement
terms \cite{DeGrand:1999gp}.  Perturbative
renormalizations are small for smeared links and tree-level estimates,
or the mean-field improved estimates used here, are sufficient to
remove $\calO (a)$ errors in the lattice spacing $a$, to all orders in
the strong coupling $g$.  The key is that both the energy
dimension-five Wilson term and the Clover term
\cite{Sheikholeslami:1985ij} are constructed with smooth links, while
the relevant operators, surviving in the continuum limit, are
constructed with the original untouched links generated via standard
Monte Carlo techniques.  We call this action the
Fat-Link-Irrelevant-Clover (FLIC) fermion action.

The established approach to nonperturbative (NP) 
improvement \cite{Luscher:1996sc} tunes the coefficient of the clover
operator to all powers in $g^2$.  Unfortunately, this formulation of
the clover action is susceptible to the problem of exceptional
configurations as the quark mass becomes small.  Chiral symmetry
breaking in the clover fermion action introduces an additive mass
renormalization into the Dirac operator that can give rise to
singularities in quark propagators at small quark masses.  In
practice, this prevents the simulation of small quark masses and the
use of coarse lattices ($\beta < 5.7 \sim a > 0.18$~fm)
\cite{DeGrand:1999gp,Bardeen:1998gv}.

The clover term of the fermion action
requires a lattice determination of the QCD field strength tensor
$F_{\mu\nu}$.  The so-called clover version of $F_{\mu\nu}(x)$
involves the product of links around the four plaquettes centered at
$x$ in the $\mu$-$\nu$ plane.  This simple formulation is commonly
used in clover actions, but it is now known to have large ${\cal
O}(a^2)$ errors.  These errors lead to $10 \%$ errors in the topological
charge even on very smooth configurations \cite{Bonnet:2000dc}.

A key feature of FLIC fermions is that the construction of irrelevant
operators using smoothed links facilitates the use of highly improved
definitions of the QCD field strength tensor $F_{\mu\nu}$.
In particular, we employ an ${\cal O}(a^4)$ improved definition of
$F_{\mu\nu}$ \cite{sbilson} in which the standard clover-sum of four
$1 \times 1$ Wilson loops is combined with $1 \times 2$, $1 \times 3$,
$2 \times 2$ and $3 \times 3$ Wilson-loop clovers.


\begin{figure}[t]
\begin{center}
{\includegraphics[height=0.95\hsize,angle=90]{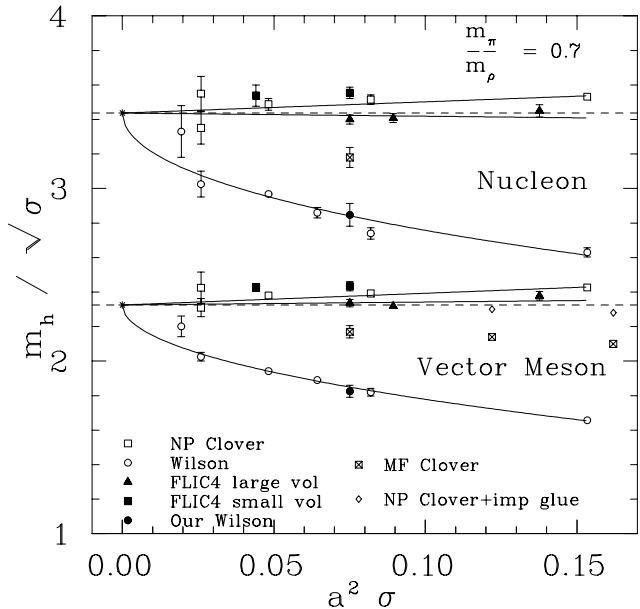}}
\vspace*{-0.0cm}
\caption{Nucleon and vector meson masses for the Wilson, Mean-Field
  (MF) improved, NP-improved clover and FLIC actions obtained by
  interpolating simulation results to $m_\pi / m_\rho = 0.7$.  For the
  FLIC action (``FLIC4''), fat links are constructed with $n=4$
  APE-smearing sweeps with smearing fraction $\alpha = 0.7$.  Results
  from the CSSM lattice collaboration are indicated by the solid
  symbols; those from earlier simulations by open or hatched symbols.
  The solid-lines illustrate fits, constrained to have a common
  continuum limit, to FLIC, NP-improved clover and Wilson fermion
  action results obtained on physically large lattice volumes.
  \vspace*{-0.3cm}
\label{scaling1} }
\end{center}
\end{figure}

The scaling analysis of FLIC fermions is performed at three different
lattice spacings and two different volumes.  The tree-level
${\calO}(a^2)$--Symanzik-improved \cite{Symanzik:1983dc} gauge action
is used on $12^3 \times 24$ and $16^3 \times 32$ lattices with lattice
spacings of 0.093, 0.122 and 0.165~fm determined from a string tension
analysis incorporating the lattice Coulomb term \cite{Edwards:1997xf}.
Where necessary, we take $\sqrt\sigma=440$~MeV.  A total of 200
configurations are used in the scaling analysis at each lattice
spacing and volume.  In addition, for the light quark simulations, 94
configurations are used on a $20^3 \times 40$ lattice with
$a=0.134$~fm.  The error analysis is performed by a third-order,
single-elimination jackknife, with the $\chi^2$ per degree of freedom
obtained via covariance matrix fits.  A fixed boundary condition and
smeared sources \cite{Gusken:qx} are used for the fermions.

Figure~\ref{scaling1} displays our most comprehensive scaling analysis
to date.  The present results for the Wilson action agree with those
of Ref.~\cite{Edwards:1998nh}.  The FLIC action performs
systematically better than the mean-field improved clover action and
competes well with those obtained with the NP-improved clover fermion
action \cite{Edwards:1998nh}.

Our two different volumes used at $a^2 \sigma \sim 0.075$ indicate a
finite volume effect, which increases the mass for the smaller volumes
at $a^2 \sigma \sim 0.075$ and $\sim 0.045$.  Examination of points from
the small and large volumes separately indicates continued scaling
toward the continuum limit.  While the finite volume effect will
produce a different continuum limit value, the slope of the points
from the smaller and larger volumes agree.

Focusing on simulation results from physical volumes with extents
$\sim 2$ fm and larger, we perform a simultaneous fit of the FLIC,
NP-improved clover and Wilson fermion action results.  The fits are
constrained to have a common continuum limit and assume errors are
$\calO (a^2)$ for FLIC and NP-improved clover actions and $\calO (a)$
for the Wilson action.  An acceptable $\chi^2$ per degree of freedom
is obtained for both the nucleon and $\rho$-meson fits.  These results
indicate that FLIC fermions provide a new form of nonperturbative
${\cal O}(a)$ improvement.  The FLIC fermion results display nearly
perfect scaling indicating $\calO (a^2)$ errors are small for this
action.

\section{Baryon Resonances}

Lattice studies of baryon excitations
\cite{LEIN1,DEREK,RICHARDS,DWF,NSTAR} provide valuable insight into
the forces of confinement and the nature of QCD in the nonperturbative
regime.  They complement the high precision measurements of the $N^*$
spectrum under way at Jefferson Lab.
The simulations presented here are performed on 392
${\calO}(a^2)$--Symanzik-improved \cite{Symanzik:1983dc}
configurations of size $16^3\times 32$ at $\beta=4.60$ providing a
lattice spacing of $a = 0.122(2)$~fm.  FLIC fermions are implemented
with 4 sweeps of APE smearing at $\alpha=0.7$.

\subsection{Spin-1/2 Baryon Resonances}
\label{Nstar}

There are two nucleon interpolating fields commonly considered in
exciting the nucleon from the vacuum.  
The standard interpolating field couples a
$u$-$d$ quark pair to a scalar diquark and is $\calO (1)$ in a
nonrelativistic reduction.  The alternate interpolator is $\calO(
p^2/E^2 )$ in a nonrelativistic reduction placing two quarks in
relative $P$ waves.  Odd-parity states may be projected from
correlation functions of these interpolators.  The standard
interpolator is expected to have stronger overlap with the
lowest-lying odd-parity state due to the scalar-diquark construction
of the interpolator.

\begin{figure}[t]
\begin{center}
{\includegraphics[width=\hsize]{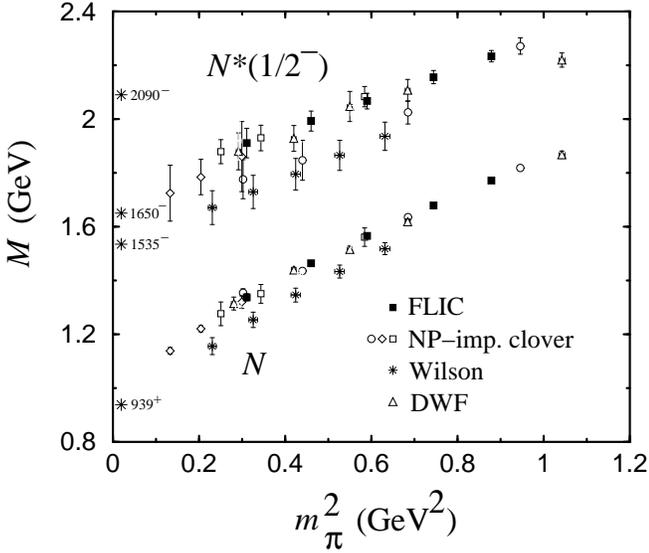}}
\vspace*{-0.4cm}
\caption{Masses of the nucleon ($N$) and the lowest $J^P={1/2}^-$
        excitation (``$N^*$''), obtained from the standard nucleon
        interpolating field.  The FLIC and Wilson results are from the
        present analysis.
\label{fig:Nstar}}
\end{center}
\vspace*{-0.3cm}
\end{figure}

In Fig.~\ref{fig:Nstar} we show the $N$ and $N^*({1/ 2}^-)$ masses
from FLIC fermions as a function of $m_\pi^2$.  For comparison, we
also show results from simulations with Wilson \cite{NSTAR} and domain
wall fermions (DWF) \cite{DWF}, and the NP-improved clover action
\cite{RICHARDS} with different source smearing and volumes.  There is
excellent agreement between the different improved actions for the
nucleon mass.  The Wilson results lie systematically low in comparison
to these due to large ${\cal O} (a)$ errors in this action
\cite{FATJAMES}.

A similar pattern is seen for the lowest-lying $N^*({1/ 2}^-)$
masses.  A mass splitting of approximately 400~MeV is clearly visible
between the $N$ and $N^*$ for all actions, including the Wilson.  The
trend of the $N^*({1/ 2}^-)$ data with decreasing $m_\pi$ is also
consistent with the mass of the lowest physical negative parity $N^*$
state.

\begin{figure}[t]
\begin{center}
{\includegraphics[width=\hsize]{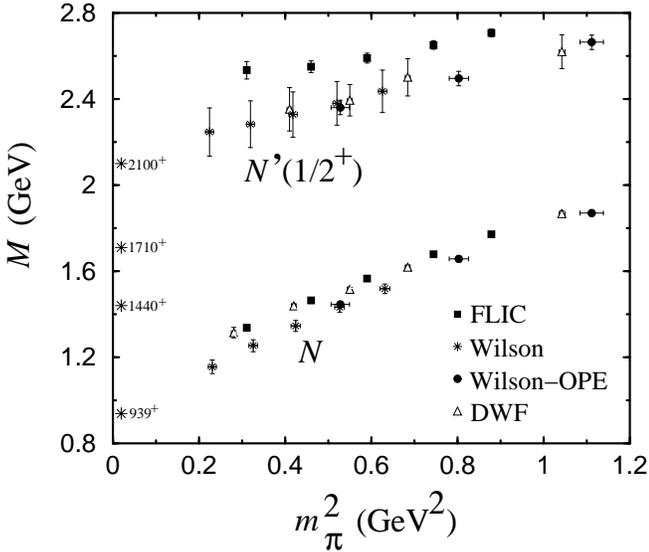}}
\vspace*{-0.4cm}
\caption{Masses of the nucleon, and the lowest $J^P={1/ 2}^+$
        excitation (``$N'$'') obtained from the alternate interpolating
        field.  The FLIC and Wilson results are from this analysis.
\label{fig:Nprime}}
\end{center}
\vspace*{-0.3cm}
\end{figure}

The mass of the $J^P = {1/ 2}^+$ state obtained from the alternate
nucleon interpolating field \cite{LEIN1}, which vanishes in a
nonrelativistic reduction, is shown in Fig.~\ref{fig:Nprime}.  In
addition to the FLIC and Wilson results from the present analysis,
also shown are the DWF results \cite{DWF}, and results from an earlier
analysis with Wilson fermions analyzed via the operator product
expansion \cite{LEIN1}.  The most striking feature of the data is the
relatively large excitation energy of the $N'$, some 1~GeV above the
nucleon.  It has been speculated that the alternate interpolator may
have overlap with the lowest ${1/ 2}^+$ excited state \cite{DWF}.
However, there is little evidence that this state is the $N^*(1440)$.
It is likely that the alternate nucleon interpolator simply does not
have good overlap with either the nucleon or the Roper, but
rather a (combination of) excited ${1/ 2}^+$ state(s).

\begin{figure}[t]
\begin{center}
{\includegraphics[width=\hsize]{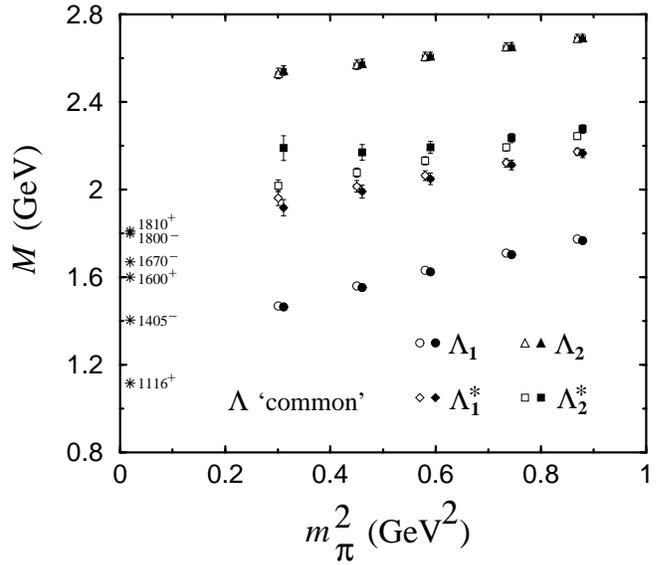}}
\vspace*{-0.4cm}
\caption{Masses of the $\Lambda({1/ 2}^\pm)$ states, obtained from the
	standard ($\Lambda_1$) and alternate ($\Lambda_2$)
	interpolating fields.  The symbols are described in the text.
\label{fig:Lambda}}
\end{center}
\vspace*{-0.3cm}
\end{figure}

The spectrum of positive and negative parity $\Lambda$ states is shown
in Fig.~\ref{fig:Lambda}.  Here we consider an interpolating field
which contains terms common to the $SU(3)$-flavor singlet and octet
interpolating fields.  It is the $SU(2)$-isospin analogue of the
$\Sigma^0$ interpolating field and does not bias the flavor symmetry
of the $\Lambda$ resonances.  The positive (negative) parity states
labeled $\Lambda_1$ ($\Lambda_1^*$) and $\Lambda_2$ ($\Lambda_2^*$)
are constructed from the traditional and alternate $\Lambda$
interpolators analogous to those of the nucleon.  The pattern of mass
splittings is similar to that observed for the $N^*$'s in
Figs.~\ref{fig:Nstar} and \ref{fig:Nprime}.  The importance of the
correlation matrix analysis projecting the eigenstates of the
Hamiltonian (filled symbols) is evident from a comparison with the
naive fits to single correlation functions (open symbols).  There is
little evidence that the $\Lambda_2$ has any significant overlap with
the first positive parity excited state, $\Lambda^*(1600)$ (c.f. the
Roper resonance, $N^*(1440)$, in Fig.~\ref{fig:Nprime}).

While it seems plausible that nonanalyticities in a chiral
extrapolation \cite{MASSEXTR} of $N_1$ and $N_1^*$ results could
eventually lead to agreement with experiment, the situation for the
$\Lambda^*(1405)$ is not as compelling.  Whereas a 150~MeV
pion-induced self energy is required for the $N_1 ,\ N_1^*$ and
$\Lambda_1$ states, 400~MeV is required to approach the empirical mass
of the $\Lambda^*(1405)$.  This large discrepancy suggests that
relevant physics may be absent from simulations in the quenched
approximation or perhaps more exotic interpolating fields are required
to obtain significant overlap with the $\Lambda^*(1405)$.
Investigations at lighter quark masses involving quenched chiral
perturbation theory will assist in resolving these issues.

\subsection{Spin-3/2 Baryon Resonances}
\label{spin32}

We consider the following isospin-${1/ 2}$, spin-${3/ 2}$ interpolator
\cite{fullSpin32}: $\chi^{N^+}_{\mu}(x) = \epsilon^{abc} \left(
u^{Ta}(x)\ C \gamma_5 \gamma_\mu\ d^b(x) \right) \gamma_5 u^c(x)$,
which transforms as a pseudo-vector under parity, in accord with a
positive parity Rarita-Schwinger spinor.  
%
%
For the $\Delta^{++}$ resonance we use the standard interpolator as in
Ref.~\cite{CHUNG}.  Since the spin-${3/ 2}$ Rarita-Schwinger
spinor-vector is a tensor product of a spin-1 vector and a spinor, the
spin-${3/ 2}$ interpolating field contains spin-${1/ 2}$
contributions.  To project a spin-${3/ 2}$ state one needs to use a
spin-${3/ 2}$ projection operator \cite{BDM}.  Following spin
projection, the correlation function for a given spin still contains
positive and negative parity states.  In an analogous procedure to
that used for spin-1/2 resonances with a fixed boundary condition in
the time direction, positive and negative parity states are obtained
by taking the trace of the correlation function with the
parity-projection operators $\Gamma_{\pm} = \left( 1\pm \gamma_4
\right)/2$.

\begin{figure}[t]
\begin{center}
{\includegraphics[height=0.95\hsize,angle=90]{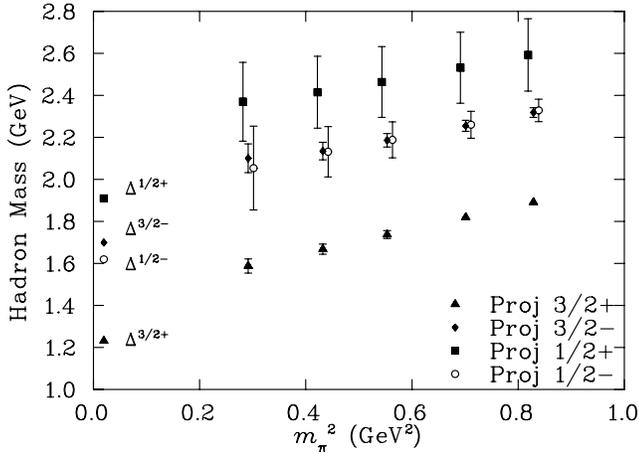}}
\vspace*{-0.0cm}
\caption{Masses of the spin-parity-projected $\Delta({3/ 2}^\pm)$
  and $\Delta({1/ 2}^\pm)$ states.  The empirical masses are
  indicated along the ordinate. 
\vspace*{-0.3cm}
\label{fig:Delta}}
\end{center}
\end{figure}

In Fig.~\ref{fig:Delta} the spin-parity-projected $\Delta({3/ 2}^+)$
(triangles) and $\Delta({3/ 2}^-)$ (diamonds) masses are shown.  We
find a clear signal for the $P$-wave $\Delta({3/ 2}^-)$ parity partner
of the $\Delta$ ground state.  The mass of the $\Delta({3/ 2}^-)$ lies
some 500~MeV above that of its parity partner.  A discernible signal
is detected for the $\Delta({1/ 2}^\pm)$ states.  The level ordering
of the $\Delta$ states is consistent with that observed in the
empirical mass spectrum.

In the isospin-${1/ 2}$ sector, large statistical fluctuations make it
difficult to obtain a clear signal, even with 392 configurations.
Parity projecting to extract the $N({3/ 2}^+)$ state, we find that the
correlation function changes sign and has a large negative
contribution for time slices in the range $t=7$--11 following the
source at $t=3$.  This behavior is an artifact associated with the
quenched decay of the excited state into $N + \eta'$ and is further
explored in Ref.~\cite{fullSpin32}.

The $N({1/ 2}^+)$ channel also displays the interplay of a quenched
decay channel and the ground state contribution.  A strong $P$-wave
coupling of the $N^*({1/ 2}^+)$ to $N \eta'$ forces the correlation
function to be negative at small times, which then turns positive at
larger times when the ground state nucleon begins to dominate the
correlation function.  This suggests that the first excited state of
the nucleon has a strong coupling to the quenched $\eta'$ which
remains degenerate with the pion in the quenched approximation.  These
results imply a gluon-rich structure for the Roper resonance and
further indicate that it may be impossible to directly observe the
Roper resonance at light quark masses in the quenched approximation.
While there are claims to have observed the Roper in quenched QCD
\cite{KENTUCKY}, these results follow from a Bayesian analysis, and
the credibility of the Bayesian-prior information used in the analysis
requires further examination \cite{Leinweber:1994gt,ALLTON}.

\begin{figure}[t]
\begin{center}
{\includegraphics[height=0.95\hsize,angle=90]{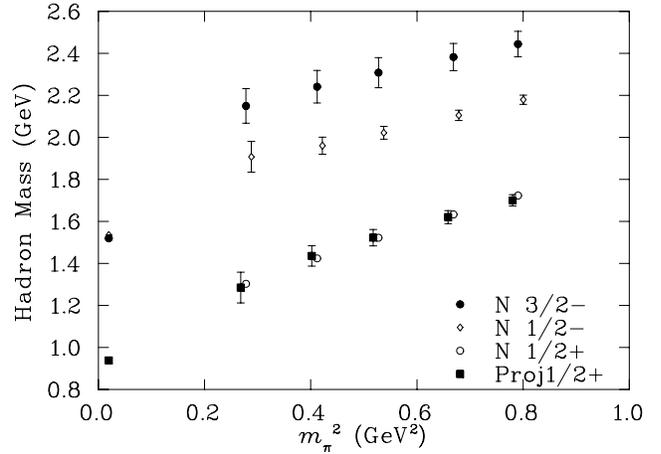}}
\vspace*{-0.0cm}
\caption{Masses of the spin-projected $N({3/ 2}^-)$ and
        $N({1/ 2}^+)$ states, compared with
        the nucleon and $N({1/ 2}^-)$ masses from
        Sec.~\protect\ref{Nstar}.
\label{fig:N32}
\vspace*{-0.3cm}}
\end{center}
\end{figure}

The extracted masses of the $N({3/ 2}^-)$ and $N({1/ 2}^+)$ states are
displayed in Fig.~\ref{fig:N32}.  Earlier results from
Sec.~\ref{Nstar} using the standard spin-${1/ 2}$ interpolating field
are also shown in Fig.~\ref{fig:N32} for reference.  There is
excellent agreement between the spin-projected ${1/ 2}^+$ state
obtained from the spin-${3/ 2}$ interpolating field and the earlier
${1/ 2}^+$ results.  

\section{Chiral Nonanalytic Behavior}
\label{chiLimit}

The FLIC fermion action has extremely impressive convergence rates for
matrix inversion \cite{FATJAMES,WASEEM}, and this provides great
promise for performing cost-effective simulations at quark masses
closer to the physical values.  Problems with exceptional
configurations have prevented such simulations with Wilson-type
fermion actions in the past.

In the absence of exceptional configurations, the standard deviation
of an observable will be independent of the number of configurations
considered in the average.  Exceptional configurations reveal
themselves by introducing a significant jump in the standard deviation
as the configuration is introduced into the average.  In severe cases,
exceptional configurations can lead to divergences in correlation
functions or prevent the matrix inversion process from converging.

The ease with which one can invert the fermion matrix using FLIC
fermions leads us to attempt simulations down to small quark masses
corresponding to $m_{\pi} / m_{\rho} = $ 0.35. 
%
%
The simulations are on a $20^3 \times 40$ lattice with a physical
length of 2.7~fm. We have used an initial set of 100 configurations,
using $n=6$ sweeps of APE-smearing and a five-loop improved lattice
field-strength tensor.  Preliminary results indicate exceptional
configurations at the few percent level \cite{fullSpin32}.  For the
current results, these configurations have been identified and removed
from the ensemble.

\begin{figure}[t]
\begin{center}
{\includegraphics[height=0.95\hsize,angle=90]{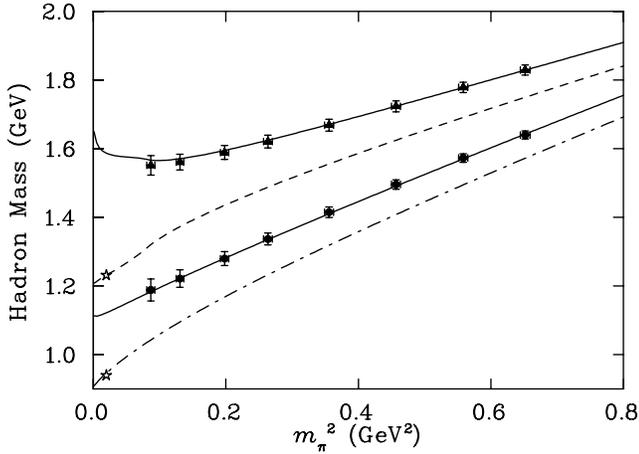}}
\vspace*{-0.0cm}
\caption{Nucleon and $\Delta$ masses for the FLIC-fermion action on a
  $20^3 \times 40$ lattice with $a=0.134$~fm.  The solid curves
  illustrate fits of finite-range regulated quenched chiral
  perturbation theory to the lattice QCD results.  The dot-dashed and
  dashed curves estimate the correction that will arise in unquenching
  the lattice QCD simulations for the $N$ and $\Delta$ respectively.
  Stars at the physical pion mass denote experimental values.
\vspace*{-0.3cm} }
\label{LQM}
\end{center}
\end{figure}

\begin{figure}[t]
\begin{center}
{\includegraphics[height=0.95\hsize,angle=90]{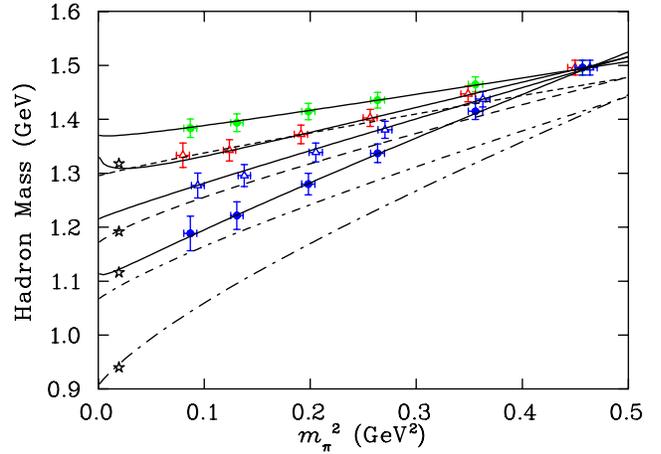}}
\vspace*{-0.0cm}
\caption{Octet baryon masses for the FLIC-fermion action.  Points are
  off set for clarity.  Symbols are as described for
  Fig.~\protect\ref{LQM}.
\vspace*{-0.3cm} }
\label{octet}
\end{center}
\end{figure}

Figure~\ref{LQM} shows the $N$ and $\Delta$ masses as a function of
$m_{\pi}^2$ for eight quark masses.  An upward curvature in the
$\Delta$ mass for decreasing quark mass is observed in the FLIC
fermion results.  This behavior, increasing the quenched $N-\Delta$
mass spitting, was predicted by Young {\em et al.} \cite{ROSS} using
quenched chiral perturbation theory (Q$\chi$PT) formulated with a
finite-range regulator.  This Q$\chi$PT fit to the FLIC-fermion results
is illustrated by the solid curves.  The dashed curves estimate the
correction that will arise in unquenching the lattice QCD simulations.
A similar preliminary analysis incorporating the light-meson cloud of
the baryon octet is illustrated in Fig.~\ref{octet}.  Inclusion of the
kaon-cloud is in progress.

\section{Conclusions}
\label{conclusion}

In constructing the irrelevant operators of clover fermion actions
with APE-smeared links, FLIC fermions provide a new form of
nonperturbative ${\cal O}(a)$ improvement.  The technique allows the
use of highly improved operators, provides optimal scaling and reduces
the exceptional configuration problem.  Quenched simulations at quark
masses down to $m_{\pi}/m_{\rho}=0.35$ have been successfully
performed on a $20^3 \times 40$ lattice with a lattice spacing of
0.134~fm.  Simulations at such light quark masses have already
revealed the non-analytic behavior of quenched chiral perturbation
theory in the $\Delta$-baryon mass.  We expect to see more evidence of
chiral nonanalytic behavior in forthcoming simulations of the
electromagnetic form factors of hadrons.

\vspace*{0.3cm}

We thank the Australian National Computing Facility for Lattice Gauge
Theory, and the Australian Partnership for Advanced Computing (APAC)
for supercomputer time.  This work was supported by the Australian
Research Council.  W.M. was supported by the U.S. Department of Energy
contract \mbox{DE-AC05-84ER40150}, under which the Southeastern
Universities Research Association (SURA) operates the Thomas Jefferson
National Accelerator Facility (Jefferson Lab).



\begin{thebibliography}{40}

\bibitem{FATJAMES}
J.M.~Zanotti {\it et al.}, 
Phys. Rev. D {\bf 60}, 074507 (2002);
Nucl. Phys. Proc. Suppl. {\bf 109}, 101 (2002).

\bibitem{ape}
M.~Falcioni {\it et al.}, 
Nucl. Phys. {\bf B251}, 624 (1985).
%

\bibitem{DeGrand:1999gp}
T.~DeGrand (MILC collaboration),
Phys. Rev. D {\bf 60}, 094501 (1999).

\bibitem{Sheikholeslami:1985ij}
B.~Sheikholeslami and R.~Wohlert,
Nucl. Phys. {\bf B259} 572, (1985).

\bibitem{Luscher:1996sc}
M.~Luscher {\it et al.}, 
Nucl. Phys. {\bf B478}, 365 (1996).
%
M.~Luscher {\it et al.}, 
Nucl. Phys. {\bf B491}, 323, 344 (1997).

\bibitem{Bardeen:1998gv}
W.~Bardeen {\it et al.}, 
Phys. Rev. D {\bf 57}, 1633 (1998);
%
W.~Bardeen {\it et al.}, 
Phys. Rev. D {\bf 57}, 3890 (1998).

\bibitem{Bonnet:2000dc}
F.D.~Bonnet {\it et al.}, 
Phys. Rev. D {\bf 62}, 094509 (2000).

\bibitem{sbilson}
S.~Bilson-Thompson {\it et al.},
Nucl. Phys. Proc. Suppl. {\bf 109}, 116 (2002);
hep-lat/0203008.

\bibitem{Symanzik:1983dc}
K.~Symanzik,
Nucl. Phys. {\bf B226}, 187 (1983).

\bibitem{Edwards:1997xf}
R.~G.~Edwards {\it et al.}, 
Nucl.\ Phys.\ B {\bf 517}, 377 (1998)
[hep-lat/9711003].

\bibitem{Gusken:qx}
S.~Gusken,
Nucl. Phys. Proc. Suppl. {\bf 17}, 361 (1990).

\bibitem{Edwards:1998nh}
R.G.~Edwards, U.M.~Heller and T.R.~Klassen,
Phys. Rev. Lett. {\bf 80}, 3448 (1998);
see also R.D.~Kenway,
Nucl. Phys. Proc. Suppl. {\bf 73}, 16 (1999),
for a review.


\bibitem{LEIN1}
D.~B.~Leinweber,
Phys. Rev. D {\bf 51}, 6383 (1995).

\bibitem{DEREK}
F.~X.~Lee and D.~B.~Leinweber,
Nucl. Phys. Proc. Suppl. {\bf 73}, 258 (1999).

\bibitem{RICHARDS}
C.~R.~Allton {\it et al.},
Phys. Rev. D {\bf 47}, 5128 (1993);
D.~G.~Richards {\em et al.},
Nucl. Phys. Proc. Suppl. {\bf 109}, 89 (2002);
%
M.~G\"ockeler {\em et al.},
Phys. Lett. B {\bf 532}, 63 (2002).

\bibitem{DWF}
S.~Sasaki, T.~Blum and S.~Ohta,
Phys. Rev. D {\bf 65}, 074503 (2002).




\bibitem{NSTAR}
W.~Melnitchouk {\em et al.},
hep-lat/0202022;
%
Nucl. Phys. Proc. Suppl. {\bf 109}, 96 (2002).

\bibitem{MASSEXTR}
D.~B.~Leinweber {\em et al.},
Phys. Rev. D {\bf 61}, 074502 (2000).
 



\bibitem{fullSpin32}
J.~M.~Zanotti {\it et al.},
in preparation.

\bibitem{CHUNG}
Y.~Chung {\em et al.},
Nucl. Phys. {\bf B197}, 55 (1982).

\bibitem{BDM}
M.~Benmerrouche {\em et al.},
Phys. Rev. C {\bf 39}, 2339 (1989).

\bibitem{KENTUCKY}
F.~X.~Lee, {\it et al.}, Lattice '02,
hep-lat/0208070;
S.~J.~Dong, {\it et al.}, Lattice '02,
hep-lat/0208055.

\bibitem{Leinweber:1994gt}
D.~B.~Leinweber,
Phys.\ Rev.\ D {\bf 51} (1995) 6369.

\bibitem{ALLTON}
C.~Allton {\it et al.}, in preparation.



\bibitem{WASEEM}
W.~Kamleh {\it et al.},
Phys. Rev. D {\bf 66}, 014501 (2002).


\bibitem{ROSS}
R.D.~Young {\it et al.},
hep-lat/0205017.

\end{thebibliography}
\end{document}